\documentclass[prl,twocolumn,superscriptaddress]{revtex4}
\usepackage{graphicx}
\begin{document}
\title{The Extended Bose-Hubbard Model on a Honeycomb Lattice}
\author{ Jing Yu Gan}
\affiliation{Center for Advanced Study, Tsinghua University,
Beijing, 100084, China}
\author{Yu Chuan Wen }
\affiliation{Interdisciplinary Center of Theoretical Studies, CAS,
Beijing 100080, China} \affiliation{Institute of Theoretical
Physics, CAS, Beijing 100080, China}
\author{ Jinwu Ye}
\affiliation{Department of Physics, The Pennsylvania State
University, University Park, PA, 16802, U.S.A. }
\author{ Tao Li}
\affiliation{Department of Physics, Renmin University of China,
Beijing 100872, China}
\author{ Shi-Jie Yang}
\affiliation{Department of Physics, Beijing Normal University,
Beijing 100875, China}
\author{ Yue Yu}
\affiliation{Institute of Theoretical Physics, CAS, Beijing 100080,
China}

\begin{abstract}
We study the extended Bose-Hubbard model on a two-dimensional
honeycomb lattice by using large scale quantum Monte Carlo
simulations. We present the ground state phase diagrams for both the
hard-core case and the soft-core case. For the hard-core case, the
transition between $\rho=1/2$ solid and the superfluid is first
order and the supersolid state is unstable towards phase separation.
For the soft-core case, due to the presence of the multiple
occupation, a stable particle induced supersolid  ( SS-p ) phase
emerges when $1/2<\rho<1$. The transition from the solid at
$\rho=1/2$ to the SS-p is second order with the superfluid density
scaling as $ \rho_{s} \sim \rho-1/2 $. The SS-p has the same
diagonal order as the solid at $ \rho=1/2 $. As the chemical
potential increasing further, the SS-p will turn into a solid where
two bosons occupying each site of a sublattice through a first order
transition.  We also calculate the critical exponents of the
transition between $\rho=1/2$ solid and superfluid at the Heisenberg
point for the hard core case. We find the dynamical critical
exponent $z=0.15$, which is smaller than results obtained on smaller
lattices. This indicates that $ z $ approaches zero in the
thermodynamic limit, so the transition is also first order even at
the Heisenberg point.
\end{abstract}

\pacs{75.10.Jm, 05.30.Pr,75.10.-b}

\maketitle

%The results are qualitatively similar to those obtained for square
%lattice.

\begin{figure}[t]
\includegraphics*[width=9cm]{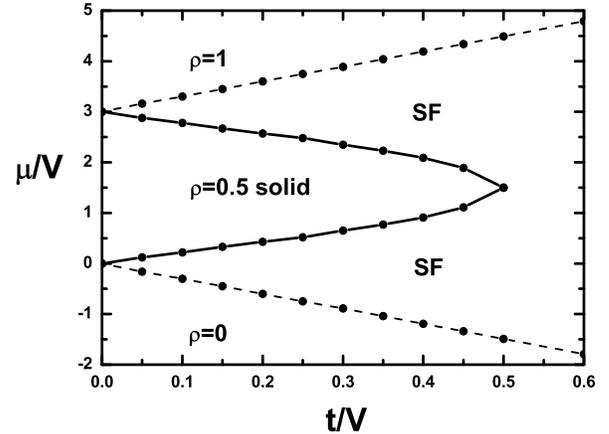}
\caption{The ground-state phase diagram for 2D hard-core
Bose-Hubbard model on the honeycomb lattice in the grand canonical
ensemble obtained from quantum Monte carlo simulation. SF:
superfluid phase.   The solid line is first order transition, while
the dashed line is second order. } \label{fig:pd}
\end{figure}

\begin{figure}[ht!!!]
\includegraphics*[width=9cm]{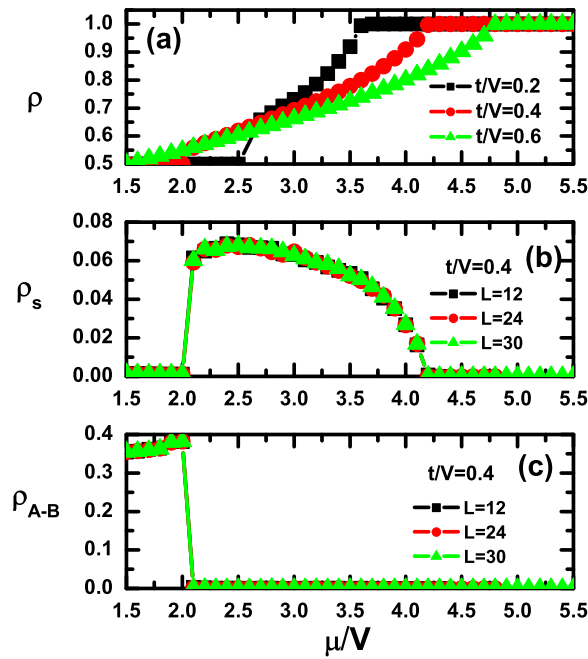}
\caption{(color online). (a) The boson density $\rho$ as a function
of chemical potential $\mu$ for $\beta=100V$ and $L=24$. (b) and (c)
are superfluid density $\rho_s$ and $\rho_{A-B}$ as functions of
$\mu$ for $t/V=0.4$, $\beta=100V$, and $L=12,24,30$ respectively.}
\label{rhohard}
\end{figure}

Bose-Hubbard model and its derivatives have been extensively studied
as the low energy effective models for ultracold atoms in optical
lattices where the superfluid to the Mott insulator transition
exists in the zero temperature \cite{bloch, fisher, jaksch}, as well
as models for the possible supersolid phases in both bipartite and
frustrated  lattices \cite{1dss, batouni, sengupta, wessel,
heidarian, melko, boninsegni, five, univ, melkostripetri, gan,
isakov, tri2e}. Similar to the Fermi Hubbard model, the Bose-Hubbard
model can be used to explore the effects of the strong correlations.
Unlike the fermi system, there is no "sign problem" in bosonic case,
so the Quantum Monte Carlo (QMC) simulations can be successfully
performed. Many interesting results have been found in these models,
such as supersolid phase, valence-bond-solid phase and striped
phase.

In the past several years, although the extended Bose-Hubbard model
on two-dimensional square lattice~\cite{batouni, sengupta} and
triangular lattice~\cite{wessel, heidarian, melko, boninsegni, gan}
has been studied extensively using QMC methods, another common
lattice, honeycomb lattice has rarely been studied by QMC. In
\cite{univ}, by using a dual vortex method (DVM), one of the authors
studied phases such as superfluid, solid, supersolid and quantum
phase transitions in an
    extended boson Hubbard model such as Eqn.\ref{boson}
    slightly away from half filling on bipartite lattices such as honeycomb and square lattice.
    He found that in the insulating side, different kinds of supersolids are generic
    possible states slightly away from half filling and also proposed a new kind of supersolid: valence bond
    supersolid which maybe stabilized by possible ring exchange interactions.
    He showed that the quantum phase transitions from solids to
    supersolids driven by a chemical potential are in the
    same universality class as that from a Mott insulator to a superfluid,
    therefore have exact exponents $ z=2, \nu=1/2, \eta=0 $.
    For example, near the solid to the supersolid transition, the superfluid
    density inside the supersolid phase  should scale as $ \rho_{s} \sim |\rho-1/2|^{(d+z-2)\nu }=|
    \rho-1/2|= \delta f $ with possible logarithmic corrections.
    Comparisons with previous quantum Monte-Carlo (QMC) simulations on a square lattice are made.

     The DVM is a magnetic space group ( MSG )  symmetry-based approach which can be used to classify
      some important phases
        and phase transitions. But the question if a particular phase will appear or
        not as a ground state can not be addressed in this
        approach, because it depends on the specific values of all the
        possible parameters in the EBHM in Eqn.\ref{boson}.
        So a microscopic approach such as Quantum Monte-Carlo (QMC)
        is needed to compare with the DVM.
        The DVM can guide the QMC to search for particular phases
        and phase transitions in a specific model. Finite size
        scalings in QMC can be used to confirm the universality
        class discovered by the dual approach.
    So far, there is no QMC simulations on a honeycomb
    lattice. In this paper, we study the extended Bose-Hubbard model on
    a honeycomb lattice by QMC simulations and also compare with the
    results achieved in \cite{univ} by the DVM. We find that the ground-state phase diagram is
qualitatively similar to the one obtained on square lattice. In the
hard-core limit, the supersolid state is unstable towards the phase
separation. The transition between $\rho=1/2$ solid and the
superfluid is the first order one. In the soft-core case, due to the
presence of the multiple occupation, a stable particle doped
supersolid ( SS-p ) phase emerges when bosons are doped into
$\rho=1/2$ solid (i.e., fillings $\rho>1/2$). We find the solid at $
\rho=1/2 $ to the SS-p at $ \rho > 1/2 $ is a second order
transition, the superfluid density inside the SS-p scales as $
\rho_{s} \sim \rho-1/2 $ which is consistent with the result
achieved in \cite{univ} by the DVM. Very precise finite size
scalings by QMC \cite{un} are underway to test the exact exponents $
z=2, \nu=1/2, \eta=0 $. However, a hole doped supersolid phase
remains unstable to phase separation when vacancies are doped into
$\rho=1/2$ solid (i.e., fillings $\rho < 1/2$). We also calculate
the critical exponents of the transition between $\rho=1/2$ solid
and superfluid at the Heisenberg point. We find the dynamical
critical exponent $z$ and the correlation length exponent $\nu$ are
$z=0.15$ and $\nu=0.38$. The dynamical critical exponent we have
obtained is smaller than it was previously obtained \cite{hebert}.
We speculate that  $ z $ will approach zero as the size of the
simulated system goes to infinity, so the the SF to the solid
transition at the Heisenberg point remains first order.

The extended Bose-Hubbard model we study can be written as:
\begin{eqnarray}
H & = &-t\sum_{<ij>}(a_{i}^{\dagger }a_{j}+a_{j}^{\dagger }a_{i})
+\frac{U}{2}\sum_{i}n_{i}(n_{i}-1) \nonumber\\
&& +V\sum_{<ij>}n_{i}n_{j} - \mu\sum_i n_i
\label{boson}
\end{eqnarray}
where $a_i^{\dagger } (a_i)$ is the creation (annihilation) operator
of the bosonic atom at site $i$; $n_i=a_{i}^{\dagger}a_i$ is the
occupation number and $\mu$ is the chemical potential. $<ij>$ runs
over nearest neighbors. $U$ and $V$ represent the on-site and
nearest-neighbor repulsive interaction, respectively. When
$U/t\rightarrow \infty$, the Hamiltonian reduces to the hard-core
one with no multiple occupation. The method we employ is  the
stochastic series expansion Quantum Monte Carlo (QMC) method with
directed loop (SSE)~\cite{sandvik}.

In order to characterize different phases, usually one studies the
static structure factor $S(Q)$ and superfluid density
$\rho_s$~\cite{pollock},
\begin{eqnarray}
S(Q) & = & \frac{1}{N}\sum_{ij}e^{iQ\cdot(r_i - r_j)} <n_i n_j>,
\nonumber\\
\rho_s & = & \frac{\langle W^2\rangle}{4\beta t},
\end{eqnarray}
where $W$ is the winding number fluctuation of the bosonic lines,
$\beta$ is the inverse temperature, and $N=2\times L\times L$ is the
lattice size. In a honeycomb lattice, $S(Q=(4\pi/3,0))$ is always
nonzero when $\rho>0$, because $Q=(4\pi/3,0)$ is not only the wave
vector for the solid phase but also the reciprocal vector on a
honeycomb lattice. So it can not be used to characterize the solid
order on a honeycomb lattice. Here we use the density difference
between the two sublattice, $\rho_{A-B}$, to characterize the solid
order for a honeycomb lattice,
\begin{eqnarray} \rho_{A-B} & = &
|\rho_A - \rho_B|,
\end{eqnarray}
where $\rho_A$ and $\rho_B$ are the boson density for $A$ and $B$
sublattice respectively. The two parameters $\rho_s$ and
$\rho_{A-B}$ can characterize different phases in the system: a
solid phase is characterized by $\rho_{A-B}\neq 0$ and $\rho_s=0$;
a SF phase by $\rho_{A-B}=0$ and $\rho_s\neq 0$; a SS phase by
both $\rho_{A-B}\neq 0$ and $\rho_s\neq 0$. Different solid phases
can be categorized by different values of $\rho_{A-B}$.

We first consider the hard-core limit, which is relatively simple.
The phase diagram is presented in Fig. 1. This phase diagram can be
qualitatively explained by strong-coupling arguments. For $\mu<-3t$,
the system is empty, while for $\mu>3(t+V)$, it is a Mott insulator
with one boson per site. At large values of $t/V$, the system
exhibits a superfluid phase. A solid phase with $\rho=1/2$ emerges
in small values of $t/V$. At $\mu/V=1.5$, the critical point between
superfluid phase and $\rho=1/2$ solid is at $(t/V)_C=0.5$. This
point $(t/V,\mu/V)=(0.5,1.5)$ is the Heisenberg point.

When bosons (holes) are doped into the $\rho=1/2$ solid, the domain
wall proliferation mechanism, i.e., the additional bosons (holes)
can hop freely across the domain wall, gain the kinetic energy
linearly in $t$, hence excludes the stable supersolid
state~\cite{wessel,sengupta}. The transition from the $\rho=1/2$
solid to the superfluid is the first order one (see Fig. 2(a)). The
phase diagram is qualitatively similar to the one obtained for a
square lattice. This is not surprising since both the square lattice
and the honeycomb lattice are bipartite lattices. However, the phase
boundary is quantitatively different. This comes from the different
coordinate number of two lattices, i.e., $q=3$ for the honeycomb
lattice and $q=4$ for the square lattice. Fig. 2 (a) shows the
density $\rho$ as a function of chemical potential $\mu$. For small
values of $t/V$ (say $t/V=0.2,0.4$), there is a jump from $\rho=1/2$
to $\rho>1/2$, which indicates a first-order transition. In the
grand canonical ensemble, a jump in $\rho$ is the token of a PS
region. Fig. 2 (b) and (c) are the superfluid $\rho_s$ and
$\rho_{A-B}$ as functions of $\mu$ for $t/V=0.4$. For $\mu/V\leq 2$,
$\rho_s=0$ while $\rho_{A-B}\neq 0$, the ground-state is the
$\rho=1/2$ solid; for $2\leq\mu/V\leq 4.2$, $\rho_s\neq 0$ while
$\rho_{A-B}=0$, the ground-state is a superfluid phase; at the
critical point $\mu_C/V=2$, both $\rho_s$ and $\rho_{A-B}$ have a
jump, indicating the first order transition; for $\mu/V>4.2$, both
$\rho_s=0$ and $\rho_{A-B}=0$ with $\rho=1$, the ground-state is a
Mott insulator. There is no region for both $\rho_s$ and
$\rho_{A-B}$ nonzero, i.e., no stable supersolid state.

\begin{figure}
\includegraphics*[width=9cm]{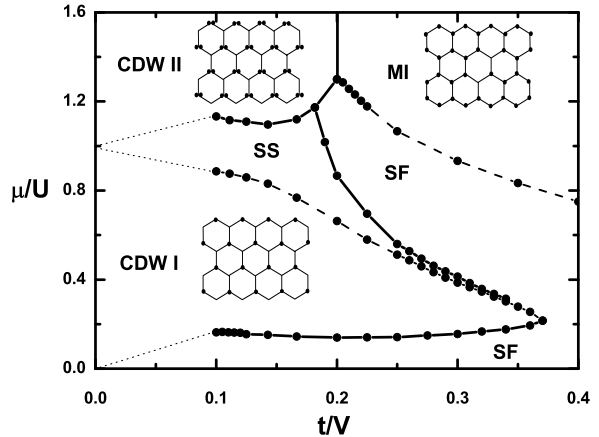}
\caption{The ground-state phase diagram for 2D soft-core
Bose-Hubbard model on the honeycomb lattice obtained from quantum
Monte carlo simulation. SF: superfluid phase; SS: supersolid phase;
PS: phase separation; CDW I: $\rho=1/2$ solid with one sublattice
occupied and one boson per site; CDW II: $\rho=1$ solid with one
sublattice occupied and two bosons per site; MI: Mott insulator
phase with one boson per site.   The solid line is first order. The
dashed line is second order. } \label{fig:pd}
\end{figure}

In the above we have found that there is no stable supersolid phase
in the hard-core Bose-Hubbard model in a honeycomb lattice, similar
to a square lattice. It was found that by softening the onsite
interaction $ U $, i.e., in the soft-core case, a stable supersolid
state emerges on the 2D square lattice~\cite{sengupta}. Motivated by
this, we next consider the soft-core case on the 2D honeycomb
lattice. Hereafter we fix $U=15t$ and $\beta=2L$ in our
calculations.

Fig. 3 shows the phase diagram for the soft-core honeycomb lattice
at $U=15t$. The phase diagram is not symmetric respect to
$\mu/V=1.5$ because finite $U$ breaks the particle-hole symmetry.
Besides the $\rho=0$ solid, $\rho=1/2$ solid (we named it CDW I
here) and the Mott insulator ($U>3V$) which already emerge in the
hard-core case, another $\rho=1$ solid ($U<3V$) with one sublattice
occupied by two bosons per site emerges. We call this phase CDW II.
A supersolid state emerges when atoms are doped into the $\rho=1/2$
solid, while the supersolid phase is unstable towards the phase
separation when holes are doped into the $\rho=1/2$ solid due to the
domain wall mechanism just like what happens in the hard-core case.
The main reason for the emergence of the CDW II and supersolid
state, is the possible multiple occupation in the soft-core case.
When bosons are doped into the $\rho=1/2$ solid, the additional
bosons can occupy either occupied or unoccupied sites. When
$|3V-U|\sim t$, the doped bosons can form a superfluid on top of the
$\rho=1/2$ solid background, this phase is a supersolid phase.
Similar results have been found in the 2D soft-core square
lattice~\cite{sengupta}, 2D triangular lattice~\cite{gan} and 1D
chain~\cite{batrouni1d}.

\begin{figure}[ht!!!]
\includegraphics*[width=9cm]{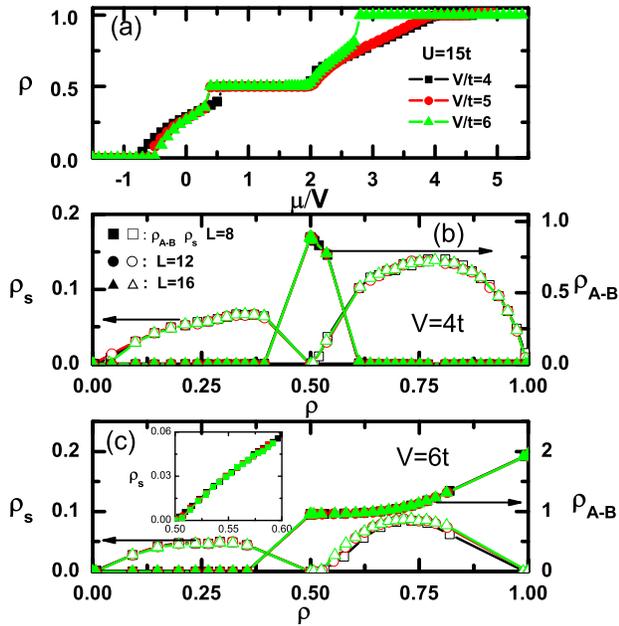}
\caption{(color online). (a): The boson density $\rho$ as a function
of $\mu$ for $V=4t,5t,6t$ at $U=15t$ and $\beta=2L$ for soft-core
case. (b) (c): The finite size scaling of superfluid density
$\rho_s$ and diagonal long range order $\rho_{A-B}$ as a function of
$\rho$ for $V=4t,6t$ respectively for soft-core case. }
\label{scaling_soft}
\end{figure}

Fig. 4 (a) is the plot of $\rho$ v.s. $\mu$ for $U=15t$.For a
density $\rho$ just less than $1/2$, there are jumps in $\rho-\mu$
plot, which indicate the transition is a first order transition and
the ground-state is a phase separation. This also indicates that the
possible hole-doped supersolid ( SS-h )  phase is unstable against
the phase separation. The jump also emerges in the fillings just
less than $1$ for large $V$ (say $V=6t$). This indicates that the
CDW II to the supersolid transition is 1st order. Fig. 4 (b) and (c)
shows the finite size scaling of superfluid density $\rho_s$ and the
diagonal long range order $\rho_{A-B}$ as functions of $\rho$ for
$V=4t$ and $6t$, respectively. From these figures, we can identify
the regions of superfluid with $\rho_s\neq 0$; $\rho=1/2$ solid with
$\rho_{A-B}\neq 0$; supersolid with both $\rho_s\neq 0$ and
$\rho_{A-B}\neq 0$; phase separation with $\rho_s$ and $\rho_{A-B}$
showing discontinuity. In the inset of Fig.\ref{scaling_soft}(c) we
also find that near $\rho=1/2$, superfluid density scales as $\rho_s
\sim (\rho-1/2)^1$.  This is consistent with the superfluid scaling
derived by the DVM in \cite{univ}. $ \rho_{A-B} $ also changes
smoothly acrose the the CDW-I to the SS-p transition. So the SS-p
has the same diagonal order as the CDW-I. These facts conclude that
the transition from CDW I to SS-p is of second order. In fact, as
shown by the DVM, the transition is in the
 same universality class as that from a Mott insulator to a superfluid,
 therefore has exact exponents $ z=2, \nu=1/2, \eta=0 $. Very precise finite size
scalings by QMC \cite{un} are underway to test the universality
class \cite{un}. Possible ring exchange terms will also be added in
\cite{un} to test the possible valence bond supersolid proposed by
the DVM in \cite{univ}.

  As shown by the DVM in \cite{univ}, due to the change of saddle
  point structure of the dual gauge field on the CDW-I and SF, the
  SF to the CDW-I is a first order transition. The SF to the SS-p is
  also first order. Fig.1 shows that the SF to the CDW-I
  is indeed first order. However, in Fig.3, our QMC data in Fig.3 can not
  determine if the SF to the SS-p is first order or second order.
  More precise QMC such as double-peak histogram maybe needed to determine the nature of the SF to
  the SS-p transition.

\begin{figure}
\includegraphics*[width=9cm]{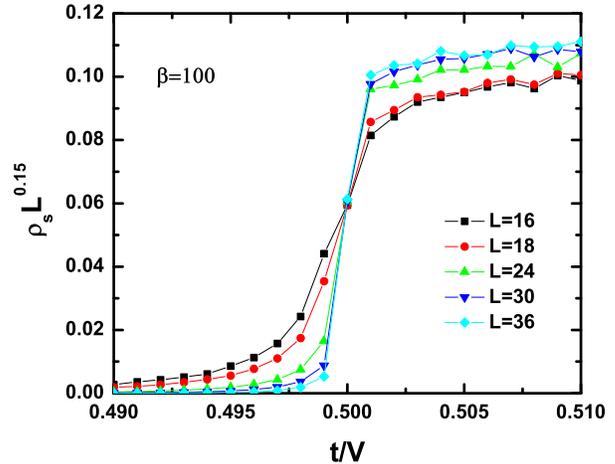}
\caption{(color online). $\rho_s$ as a function of $t/V$ for
$L=16,18,24,30,36$ for $\beta=100V$ at $\mu/V=1.5$. The intersection
at $t/V=0.5$ is the critical point.} \label{z}
\end{figure}

For the hard core case, at $t/V=0.5$, corresponding to the isotropic
Heisenberg point of the spin-1/2 quantum antiferromagnetic model,
the half filled solid melts to superfluid. The critical behavior of
this Heisenberg point is controversial. By assuming a second order
transition, H\'{e}bert {\it et.al.} \cite{hebert} found rather
exotic critical exponents, $z=0.25$ and $\nu=0.36$. However, by
studying the XXZ model crossing the isotropic point, Sandvik and
Melko point out that the critical exponent $z$ will become zero in
the thermodynamic limit \cite{sandvik-melko}.

In the below, by taking the same strategy as H\'{e}bert {\it et.al.}
\cite{hebert}, e.g. assuming it is a second order quantum phase
transition crossing the Heisenberg point, we also calculate the
critical exponents for the 2D hard-core boson model on the honeycomb
lattice at the Heisenberg point. The finite size scaling function of
superfluid density is assumed to have the following form
\cite{fisher}:
\begin{eqnarray}
\rho_s & = & \frac{1}{L^z}F((t/V-(t/V)_C)L^{1/\nu}, \beta/L^z),
\end{eqnarray}
where $z$ is the dynamical critical exponent, $\nu$ is the
correlation length exponent, and $F$ is the corresponding scaling
function. Note that the finite size scaling function $F$ does not
only depend on the scaled distance to the critical point
($(t/V-(t/V)_C)L^{1/\nu}$ ), but also on the ratio of the inverse
temperature and the lattice size ($\beta/L^z$). To obtain the
critical exponents, one should simulate several sets of lattice
sizes ($L$), each set with different values of $z$. Here we follow
the procedure of Ref.~\cite{hebert} for the square lattice, e.g. we
fix a large enough $\beta$ so that we can assume the second term of
$F$ is a constant as L is varied. With this choice, $\rho_sL^z$ must
be independent of $L$ at the critical point ($t/V=0.5$). We
calculate $\beta=20, 40, 60, 80,$ and $100$ and find the results
depend little on temperature. In the below, we present results only
of $\beta=100$.

\begin{figure}
\includegraphics*[width=9cm]{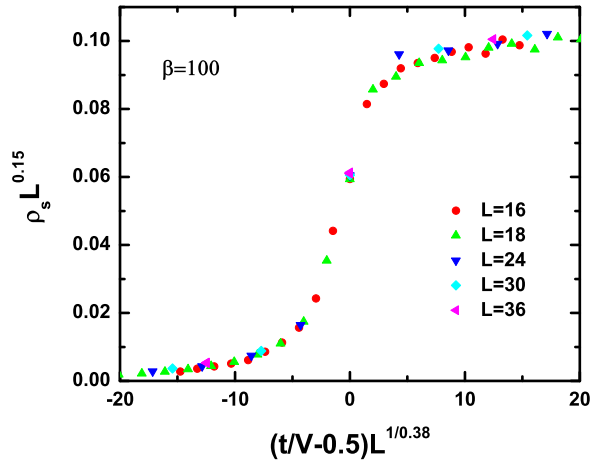}
\caption{(color online). Data collapse of the superfluid density
$\rho_s$ for $\beta=100V$ at $\mu/V=1.5$. This yields the critical
exponents $z=0.15$ and $\nu=0.38$.} \label{nu}
\end{figure}

First we simulate on smaller lattices $L=6, 8, 10, 12$ and we obtain
the same exponents as Ref.~\cite{hebert} (not show here). Then we
simulate on larger lattices, e.g. $L=16, 18, 24, 30, 36$. The best
intersection we found is at $z=0.15$ (Fig. \ref{z}). Replotting this
figure by the scaled variable $(t/V-0.5)L^{1/\nu}$, the best data
collapse we found is at $\nu=0.38$ (see fig. \ref{nu}).

The dynamical exponent we have obtained on larger lattices is
smaller than that on smaller lattices. We speculate that it will
approach zero in the thermodynamic limit, so the transition is first
order, though it maybe very weak. This scenario is consistent with
Sandvik's conclusion on XXZ model on square lattice
\cite{sandvik-melko}.

In conclusion, we have studied the extended Bose-Hubbard model on a
two-dimensional honeycomb lattice by using the quantum Monte Carlo
simulation with the stochastic series expansion (SSE). We present
the results both in the hard-core and soft-core case. We find that
for the hard-core case, the supersolid state is unstable towards the
phase separation due to the domain wall proliferation mechanism. The
transition between $\rho=1/2$ solid and the superfluid is the first
order one. For the soft-core case, due to the presence of the
multiple occupation, a stable particle induced supersolid ( SS-p )
phase emerges when atoms are doped into $\rho=1/2$ solid (i.e.,
fillings $\rho>1/2$), while for fillings $\rho<1/2$, the possible
hole induced supersolid (SS-h ) phase is unstable towards phase
separation. The results are qualitatively similar to the one
obtained for a square lattice.  We found the CDW-I to the SS-p
transition is second order with the superfluid density  inside the
SS-p scaling as $ \rho_{s} \sim \rho-1/2 $, the SS-p has the same
diagonal order as the CDW-I ( Fig.3). However, the SS-p to the
CDW-II transition is first order with jumps in both $ \rho_{s} $ and
$ \rho_{A-B} $. All these results are consistent with those achieved
by the DVM in \cite{univ}. We also calculate the critical exponents
of the transition between $\rho=1/2$ solid and the superfluid at
Heisenberg point. The dynamical critical exponent $z$ and the
correlation length exponent $\nu$ we find is $z=0.15$ and
$\nu=0.38$. Howver, w expect that $ z $ is size dependent and
approaches zero in thermodynamic limit. This fact indicates  that
the solid to the SF transition is still 1st order even at the
Heisenberg point.

{\it Note added.} Recently, we become aware of a similar work
\cite{wessel-honey}, where similar results are obtained.

\begin{acknowledgments}
Y.C.Wen and J.Y.Gan would like to thank S. M. Li, X. C. Lu, S. J.
Qin and Z. B. Su for helpful discussion. J. Ye thanks Prof.
Quangshan Tian for hospitality during his visit at Beijing
university. This work was supported in part by Chinese National
Natural Science Foundation and China Postdoctoral Science Foundation
No.20060390079. S. J. Yang is supported by NSFC under grant No.
10574012. The simulations were performed on the HP-SC45 Sigma-X
parallel computer of ITP and ICTS, CAS .
\end{acknowledgments}

\end{document}